\documentclass[twocolumn,showpacs,preprintnumbers,amsmath,amssymb]{revtex4}

\usepackage{graphicx}
\usepackage[dvips]{color}

\begin{document}


\title{Precise determination of $^6$Li cold collision parameters by radio-frequency spectroscopy on weakly bound molecules}
\author{M. Bartenstein$^1$}
\author{A. Altmeyer$^1$}
\author{S. Riedl$^1$}
\author{R. Geursen$^1$}
\author{S. Jochim$^1$}
\author{C. Chin$^1$}
\author{J. Hecker Denschlag$^1$}
\author{R. Grimm$^{1,2}$}
\affiliation{$^{1}$Institut f\"{u}r Experimentalphysik,
Universit\"{a}t Innsbruck, Technikerstra{\ss}e 25, 6020 Innsbruck, Austria\\
$^{2}$Institut f\"{u}r Quantenoptik und Quanteninformation,
\"{O}sterreichische Akademie der Wissenschaften, 6020 Innsbruck,
Austria}
\author{A. Simoni}
\author{E. Tiesinga}
\author{C. J. Williams}
\author{P. S. Julienne}
\affiliation{Atomic Physics Division, National Institute of
Standards and Technology\\100 Bureau Drive Stop 8423, Gaithersburg, Maryland 20899}

\date{\today}

\begin{abstract}
We employ radio-frequency spectroscopy on weakly bound $^6$Li$_2$
molecules to precisely determine the molecular binding energies
and the energy splittings between molecular states for different
magnetic fields. These measurements allow us to extract the
interaction parameters of ultracold $^6$Li atoms based on a
multi-channel quantum scattering model. We determine the singlet
and triplet scattering lengths to be $a_s=45.167(8)a_0$ and
$a_t=-2140(18)a_0$ (1 $a_0$ = 0.0529177 nm), and the positions of
the broad Feshbach resonances in the energetically lowest three
$s-$wave scattering channels to be 83.41(15) mT, 69.04(5) mT, and
81.12(10) mT.
\end{abstract}



\maketitle

Molecular level structure near a collision threshold uniquely
determines the scattering properties of ultracold atoms. When a
molecular state is tuned near the scattering threshold, the atomic
scattering amplitude can be resonantly altered. Magnetically tuned
Feshbach resonances \cite{feshbach} in ultracold fermionic gases
have recently led to ground-breaking observations, including the
condensation of molecules \cite{li2becinn, k2bec, li2becmit,
li2becens, huletBEC} and the studies of the crossover physics from
a molecular Bose-Einstein condensate to atomic Cooper pairs in the
Bardeen-Cooper-Schrieffer state (BEC-BCS crossover)
\cite{xoverexp, li2becens, gap}. These studies are of general
importance in physics as the ultracold Fermi gas provides a unique
model system for other strongly interacting fermionic systems
\cite{xovertheo}.

In spin mixtures of $^6$Li atoms, a broad Feshbach resonance in
the energetically lowest $s$-wave channel \cite{houbiers} allows
for precise interaction tuning. This, together with the
extraordinary stability of the system against inelastic decay
\cite{Petrov,li2becinn}, makes $^6$Li the prime candidate for
BEC-BCS crossover studies. Precise knowledge of the magnetic-field
dependent scattering properties is crucial for a quantitative
comparison of the experimental results with crossover theories. Of
particular importance is the precise value of the magnetic field
where the $s-$wave scattering diverges. At this unique point, the
strongly interacting fermionic quantum gas is expected to exhibit
universal properties \cite{universal}. Previous experiments
explored the $^6$Li resonance by measuring inelastic decay
\cite{dieckmann}, elastic collisions \cite{Ohara02,zerocrossing},
and the interaction energy \cite{bourdel1}, but could only locate
the exact resonance point to within a range between 80\,mT and
85\,mT.

An ultracold gas of weakly bound molecules is an excellent
starting point to explore the molecular energy structure near
threshold \cite{debbieRF}. Improved knowledge on the exact $^6$Li
resonance position was recently obtained in an experiment that
observed the controlled dissociation of weakly bound $^6$Li$_2$
molecules induced by magnetic field ramps \cite{mitnew}. These
measurements provided a lower bound of 82.2\,mT for the resonance
position. Studies of systematic effects suggested an upper bound
of 83.4\,mT. Within this range, however, we observe the physical
behavior of the ultracold gas still exhibits a substantial
dependence on the magnetic field \cite{gap}. In this Letter, we
apply radio-frequency (rf) spectroscopy \cite{debbieRF,ketterleRF}
on weakly bound molecules to precisely determine the interaction
parameters of cold $^6$Li atoms. Together with a multi-channel
quantum scattering model, we obtain a full characterization of the
two-body scattering properties, essential for BEC-BCS crossover
physics.

\begin{figure}
\includegraphics[width=2.75in]{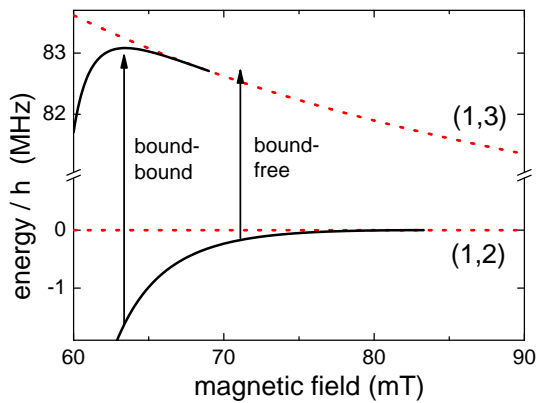}
\caption{Energy level structure near the Li$_2$ dissociation threshold as a
function of magnetic field $B$. The threshold energy of the $(1,3)$ scattering
channel (upper dotted line) is plotted relative to the $(1,2)$ threshold (lower
dotted line). In the $(1,2)$ channel, a molecular state (lower solid line)
exists below the Feshbach resonance at $\sim 83$~mT. In the $(1,3)$ channel,
another molecular state (upper solid line) exists below the resonance at $\sim
69$~mT. The bound-free and bound-bound transitions of molecules in the $(1,2)$
channel are illustrated by the arrows.} \label{fig1}
\end{figure}

The relevant atomic states are the lowest three sublevels in the
$^6$Li ground state manifold, denoted by $|1\rangle$, $|2\rangle$
and $|3\rangle$. Within the magnetic field range investigated in
this experiment, these levels form a triplet of states,
essentially differing by the orientation of the nuclear spin ($m_I
= 1, 0, -1$). Figure~\ref{fig1} shows the energy level structure
of the two scattering channels $|1\rangle+|2\rangle$ and
$|1\rangle+|3\rangle$, denoted by $(1,2)$ and $(1,3)$,
respectively. The broad Feshbach resonance occurs in the $(1,2)$
channel near 83~mT. When the magnetic field is tuned below the
resonance, atoms in the $(1,2)$ channel can form weakly bound
molecules \cite{selim1}. For the $(1,3)$ channel, a similar
Feshbach resonance \cite{ketterleRF} occurs near 69~mT.

Starting with molecules formed in the $(1,2)$ channel, we drive
the rf transition to the $(1,3)$ channel at various magnetic field
values $B$. The rf excitation can dissociate a molecule into two
free atoms (bound-free transition; see Fig.~\ref{fig1})
\cite{debbieRF} or, for $B<69$ mT, it can also drive the
transition between the molecular states in the $(1,2)$ and $(1,3)$
channels (bound-bound transition). In both processes, the rf
excitation results in loss of molecules in the $(1,2)$ channel.
This loss constitutes our experimental signal. We perform
measurements at different magnetic fields for both the bound-free
and the bound-bound transitions.

Our experimental procedure is similar to Ref.~\cite{gap}. We start
with a pure condensate of typically $2\times 10^5$ molecules at a
magnetic field of $76.4$~mT \cite{li2becinn}. The condensate is
confined in a weak optical trap, where the peak molecular density
is near $10^{13}$~cm$^{-3}$. We then linearly ramp the magnetic
field to a specific value between 66~mT and 72~mT in typically
200~ms. After the ramp, we apply a single rf pulse for 200~ms with
its frequency tuned close to the atomic transition $|2\rangle$ to
$|3\rangle$. Following the rf pulse, we apply state-selective
absorption imaging, which is sensitive to free atoms in state
$|2\rangle$ and molecules in the $(1,2)$ channel.

To precisely determine the magnetic field, we employ rf
spectroscopy on thermal atoms with temperature $T=6 T_\mathrm{F}$,
where $T_\mathrm{F}$ is the Fermi temperature. Here, the rf
transition energy corresponds to the internal energy difference
between the states $|2\rangle$ and $|3\rangle$, $h f_0$, where $h$
is Planck's constant. This energy is magnetic field dependent and
the transition frequency is about $83$~MHz in the magnetic field
range we study. The measured transition has a narrow linewidth of
less than 1 kHz, and the center position can be determined to
within a few 100~Hz. This high resolution allows us to calibrate
our magnetic field to an accuracy of a few $\mu$T based on the
Breit-Rabi formula and the $^6$Li parameters given in
\cite{breit}. Within our statistical uncertainty, we do not
observe any density-dependent frequency shifts \cite{ketterleRF}.

\begin{figure}
\includegraphics[width=2.6in]{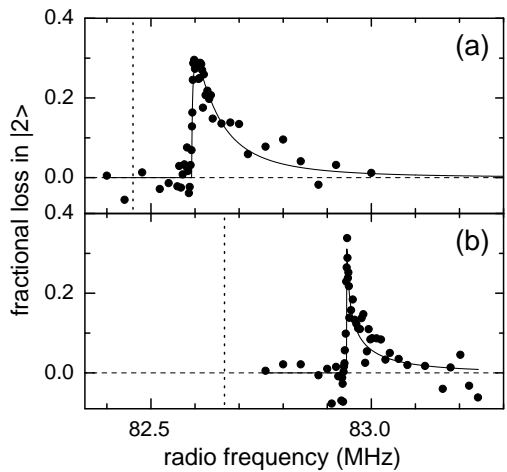}
\caption{Bound-free rf spectra at $72.013(4)$ mT (a) and $69.483(4)$ mT (b).
Fractional loss in state $|2\rangle$ is measured as a function of the radio
frequency. The solid lines are the fit based on Eq.~(\ref{eq}). The atomic
transition frequencies, which are measured independently, are indicated by the
vertical dashed lines.} \label{fig2}
\end{figure}

For bound-free transitions, the molecules in the $(1,2)$ channel
make a transition to the $(1,3)$ scattering continuum. The
excitation rate from a stationary molecule to an atomic scattering
state with kinetic energy $2E_k$ is determined by the
Franck-Condon factor between the bound and free wavefunctions
\cite{linetheory}. From energy conservation, $2E_k$ is related to
the rf transition energy $hf$ by $hf=hf_0+E_b+2E_k$, where $E_b$
is the binding energy of the molecules in the $(1,2)$ channel. The
variation of the Franck-Condon factor with atomic kinetic energy
leads to a broad and asymmetric dissociation lineshape
\cite{linetheory}.

Rf dissociation spectra at $72.0$~mT and $69.5$~mT are shown in
Fig.~\ref{fig2}. An important feature of the spectra is the sharp
rising edge on the low frequency side. This threshold corresponds
to the dissociation of a molecule into two atoms with zero
relative momentum. Therefore, the position of the edge relative to
the atomic transition directly indicates the molecular binding
energy.

We determine the dissociation threshold and thus the molecular
binding energy by fitting the full lineshape. The lineshape
function \cite{linetheory} depends on both the $(1,2)$ molecular
binding energy $E_b$ and the scattering length $a_{13}$ in the
$(1,3)$ channel. In the range of magnetic fields we investigate,
$a_{13}$ is much larger than the interaction range of the van der
Waals potential of $\sim30a_0$. The lineshape function $P(E)$ is
then well approximated by \cite{linetheory}
\begin{eqnarray}
P(E) \propto E^{-2} (E-E_b)^{1/2}(E-E_b+E')^{-1}\,, \label{eq}
\end{eqnarray}
where $E = hf-hf_0$ and $E'=\hbar^2/ma_{13}^2$. From the fits to
the experimental data \cite{fit}, we determine the threshold
positions, given in Table I. Together with the atomic transition
frequencies, we conclude that the molecular binding energies are
$E_b=h\times 134(2)$ kHz at 72.013(4) mT and $E_b=h\times 277(2)$
kHz at 69.483(4) mT.

\begin{figure}
\includegraphics[width=2.6in]{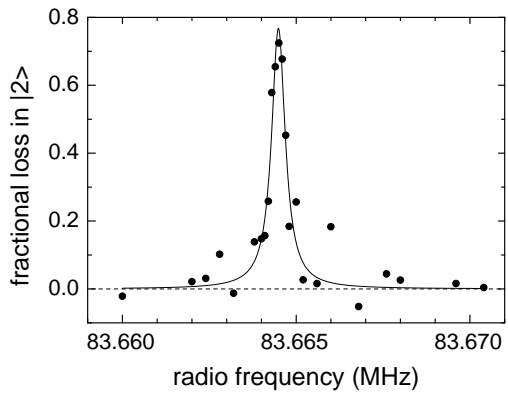}
\caption{Bound-bound rf spectrum at 66.144(2)~mT. The fractional
population loss in state $|2\rangle$ shows a narrow resonance. We
determine the center position to be 83.6645(3)~MHz from a
Lorentzian fit (solid line).} \label{fig3}
\end{figure}

For magnetic field $B<69$ mT, we can drive the rf transition
between the $(1,2)$ and $(1,3)$ molecular states. Here, the
resonance frequency is given by the energy difference of the two
molecular states.  To avoid possible systematic mean-field shifts
at these lower magnetic fields \cite{ketterleRF}, we prepare a
thermal mixture of atoms and molecules with temperature $T\approx
T_\mathrm{F}$ by a controlled heating method \cite{gap}. Rf
spectroscopy is performed at $67.6$~mT and $66.1$~mT. The
bound-bound transition signal at $66.1$~mT is shown in
Fig.~\ref{fig3}. By fitting the narrow transition line with a
Lorentzian profile, we determine the resonance frequency, see
Table I. Notably, below the resonance in the $(1,3)$ channel at
$\sim$69~mT, the bound-free transition is much weaker due to a
Fano-type interference effect \cite{linetheory}.

Because of the high precision of the measured transition
frequencies, a careful analysis of systematic effects is
necessary. Possible systematic shifts include differential light
shifts of the two molecular states and density-dependent many-body
shifts. In order to characterize these possible systematic errors,
we experimentally investigate these shifts by varying the trap
depth of the optical potential. In a deeper trap, both the
differential light shifts and mean-field shifts are expected to
increase. We repeat the bound-free and bound-bound rf spectroscopy
in traps with different laser powers $P$ between $3.8$ mW and
$310$ mW. We do not see systematic shifts within our statistical
uncertainty. The measurements show that these systematic shifts do
not exceed the uncertainties given in Table I.

Given the measured data summarized in Table I, it is possible to
predict the location of the scattering resonances in the $(1,2)$,
$(1,3)$ and $(2,3)$ channels if we have an accurate theoretical
model of the collision.  We use a standard multi-channel model for
the interaction of two  $^2$S atoms with nuclear
spin~\cite{Stoof88} to calculate the scattering lengths and bound
state energies for these channels. It is only necessary to include
$s$-waves in the basis set, since we find that there is a
negligible change within the experimental uncertainties if we also
include higher partial waves in the basis set. The interaction
potential model is the same as described in Ref.~\cite{Ohara02}.
It uses a combination of Rydberg-Klein-Rees and {\it ab initio}
potentials for the singlet ($^1\Sigma_g^+$) and triplet
($^3\Sigma_u^+$) states at short range, and joins them smoothly
onto long range potentials based on the exchange~\cite{Cote94} and
van der Waals dispersion energy~\cite{Yan96}, the lead term of
which is $C_6=1393.39(16)$ au (1 au $=9.57344 \times 10^{-26}$ J
nm$^6$). As in Ref.~\cite{Ohara02}, the singlet $^1\Sigma_g^+$ and
triplet $^3\Sigma_u^+$ scattering lengths, $a_s$ and $a_t$
respectively, are varied by making small variations to the inner
wall of the potential.  Once $a_s$ and $a_t$ are specified, all
other scattering and bound state properties for all channels of
two $^6$Li atoms are uniquely determined, including the positions
of the resonances. Consequently, varying $a_s$ and $a_t$ to fit
the binding energies and energy differences from rf spectroscopy
determines the values of these two free parameters.

Fitting the data of the present experiment determines
$a_s=45.167(8)a_0$ and $a_t=-2140(18)a_0$.  The uncertainty
includes both the uncertainty in the measured value of the
magnetic field and the uncertainty in the rf measurements. Our
scattering lengths agree within the uncertainties with previous
determinations: $a_s=45.1591(16) a_0$~\cite{mitnew} and
$a_t=-2160(250) a_0$~\cite{Abraham97}. Table I shows a comparison
of the measured and best fit calculated energies. The calculated
positions of the broad $s-$wave resonances for the $(1,2)$,
$(1,3)$, and $(2,3)$ channels are 83.41(15)~mT, 69.04(5)~mT, and
81.12(10)~mT respectively.

\begin{table}
\caption{Comparison of measured and calculated transition
frequencies. Magnetic field values in the second column are
derived from the atomic transition positions in the first column.
We report the measured peak resonance frequencies for the atomic
and molecular bound-bound transitions and the transition threshold
positions for molecular bound-free transitions. The theory values
are from the multi-channel quantum scattering calculation. Values
in parentheses indicate one $\sigma$ uncertainties.}
\begin{tabular}{cccc}\hline
  Atoms (MHz) & $B$ (mT) & Molecules (MHz) & Theory (MHz) \\ \hline
  82.96808(20) & 66.1436(20) & 83.6645(3)$^a$  & 83.6640(10) \\
  82.83184(30) & 67.6090(30) & 83.2966(5)$^a$  & 83.2973(10) \\
  82.66686(30) & 69.4826(40) & 82.9438(20)$^b$ & 82.9419(13) \\
  82.45906(30) & 72.0131(40) & 82.5928(20)$^b$ & 82.5910(13) \\ \hline
\multicolumn{4}{l}{$^a$ bound-bound transition frequency.}\\
\multicolumn{4}{l}{$^b$ bound-free transition threshold.}\\
\end{tabular}
\label{Table1}
\end{table}

\begin{figure}
\includegraphics[width=2.75in]{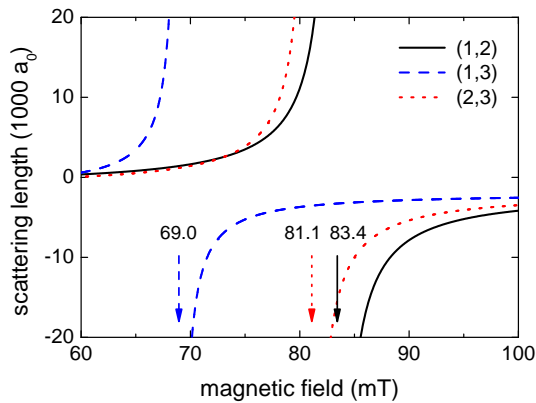}
\caption{Scattering lengths versus magnetic field from
multi-channel quantum scattering calculations for the $(1,2)$,
$(1,3)$, and $(2,3)$ scattering channels. The arrows indicate the
resonance positions.} \label{fig4}
\end{figure}

Figure~\ref{fig4} shows the scattering lengths calculated for
several different channels in the magnetic field range of interest
to BEC-BCS crossover experiments. We find that the formula
$a=a_b[1+\Delta(B-B_0)^{-1}][1+\alpha(B-B_0)]$ fits the calculated
scattering lengths to better than $99\%$ over the range of 60 to
120~mT. This expression includes the standard Feshbach resonance
term \cite{verhaar} with the background scattering length $a_b$,
resonance position $B_0$ and resonance width $\Delta$, and a
leading-order correction parameterized by $\alpha$. The respective
values for $a_b$, $B_0$, $\Delta$, and $\alpha$ are $-1405a_0$,
83.4149 mT, 30.0 mT, and 0.0040 mT$^{-1}$ for channel $(1,2)$,
$-1727a_0$, 69.043 mT, 12.23 mT, and 0.0020 mT$^{-1}$ for channel
$(1,3)$, and $-1490a_0$, 81.122 mT, 22.23 mT, and 0.00395
mT$^{-1}$ for channel $(2,3)$.

The $(1,3)$ channel molecular bound state can decay to the $(1,2)$
channel by a very weak spin-dipolar coupling.  We have used the
methods of Ref.~\cite{Koehler04} to calculate the two-body
lifetime of the $(1,3)$ bound state due to pre-dissociation to the
$(1,2)$ channel, and find that it is very long, greater than 10 s
at 60.0 mT, increasing to 1000 s at 68.5 mT very close to
resonance. However, $(1,3)$ molecules might be quenched by
collisions with $|2\rangle$ atoms or $(1,2)$ channel molecules,
since with three different spin states involved in the collision,
there would be no fermionic suppression of collision rates
according to the mechanism of Ref.~\cite{Petrov}.

In conclusion, radio-frequency spectroscopy on ultracold, weakly
bound molecules allowed us to precisely determine the molecular
binding energies and the energy splittings between two molecular
states for different magnetic fields. Based on the measured data
and a multi-channel quantum scattering model, we determine the
scattering lengths as a function of magnetic field and the
Feshbach resonance positions in the lowest three channels with
unprecedented precision. With this data, we can fully characterize
the interaction strength between particles in the BEC-BCS
crossover regime for future experiments based on $^6$Li atoms.

We acknowledge support by the Austrian Science Fund (FWF) within
SFB 15 (project part 15) and by the European Union in the
framework of the Cold Molecules TMR Network under Contract No.\
HPRN-CT-2002-00290. P. J. thanks the Office of Naval Research for
partial support. C. C.\ is a Lise-Meitner research fellow of the
FWF.


\begin{references}
\bibitem{fit}
We assume an exponential decay of the particle number and fit
$1-{\rm e}^{-\iota P(E)}$ to the loss signal, where
$\iota$ depends on the rf power. In principle, we can also
determine interaction parameters from $E'$. However, our fits
yield large uncertainties for $E'$.
\end{references}
\end{document}